\documentclass[preprint,12pt]{elsarticle}

\usepackage{amssymb}
\usepackage{hyperref}
\usepackage{amsmath}
\usepackage{color}
\usepackage[usernames,dvipsnames]{xcolor}
\RequirePackage{color}\definecolor{RED}{rgb}{1,0,0}\definecolor{BLUE}{rgb}{0,0,1} %DIF PREAMBLE

\journal{Journal of Fluids and Structures}

\begin{document}

\begin{frontmatter}



\title{Machine Learning Assisted Resistive Force Theory for Helical Structures at Low Reynolds Number}


\author[inst1]{Sangmin Lim}

\affiliation[inst1]{organization={Department of Mechanical \& Aerospace Engineering\\ University of California, Los Angeles},%Department and Organization
            addressline={420 Westwood Plaza}, 
            city={Los Angeles},
            postcode={90024}, 
            state={CA},
            country={USA}}

\author[inst2]{Charbel Habchi}
\author[inst1]{\corref{cor1}\fnref{label2}Mohammad Khalid Jawed}
\fntext[label2]{Corresponding author, E-mail : khalidjm@seas.ucla.edu}
\affiliation[inst2]{organization={IMSIA UMR EDF-CNRS-CEA 9219, Institut Polytechnique de Paris},%Department and Organization
            addressline={EDF Lab Paris-Saclay, 7 Bd Gaspard Mong}, 
            city={Palaiseau},
            postcode={91120}, 
            country={France}}

\begin{abstract}
The hydrodynamic forces on a slender rod in a fluid medium at low Reynolds number can be modeled using resistive force theories (RFTs) or slender body theories (SBTs). The former represent the forces by local drag coefficients and are computationally cheap; however, they are physically inaccurate when long-range hydrodynamic interaction is involved. The later are physically accurate but require solving integral equations and, therefore, are computationally expensive. This paper investigates RFTs in comparison with state-of-the art SBT methods. During the process, a neural network-based hydrodynamic model that -- similar to RFTs -- relies on local drag coefficients for computational efficiency was developed. However, the network is trained using data from an SBT (regularized stokeslet segments method). The $R^2$ value of the trained coefficients were $\sim 0.99$ with mean absolute error of $1.6\times10^{-2}$. The machine learning resistive force theory (MLRFT) accounts for local hydrodynamic forces distribution, the dependence on rotational and translational speeds and directions, and geometric parameters of the slender object. We show that, when classical RFT fails to accurately predict the forces, torques, and drags on slender rods under low Reynolds number flows, MLRFT exhibits good agreement with physically accurate SBT simulations. In terms of computational speed, MLRFT forgoes the need of solving an inverse problem and, therefore, requires negligible computation time in comparison with SBT. MLRFT presents a computationally inexpensive hydrodynamic model for flagellar propulsion can be used in the design and optimization of biomimetic flagellated robots and analysis of bacterial locomotion.
\end{abstract}

\begin{graphicalabstract}
\includegraphics[width=1\textwidth]{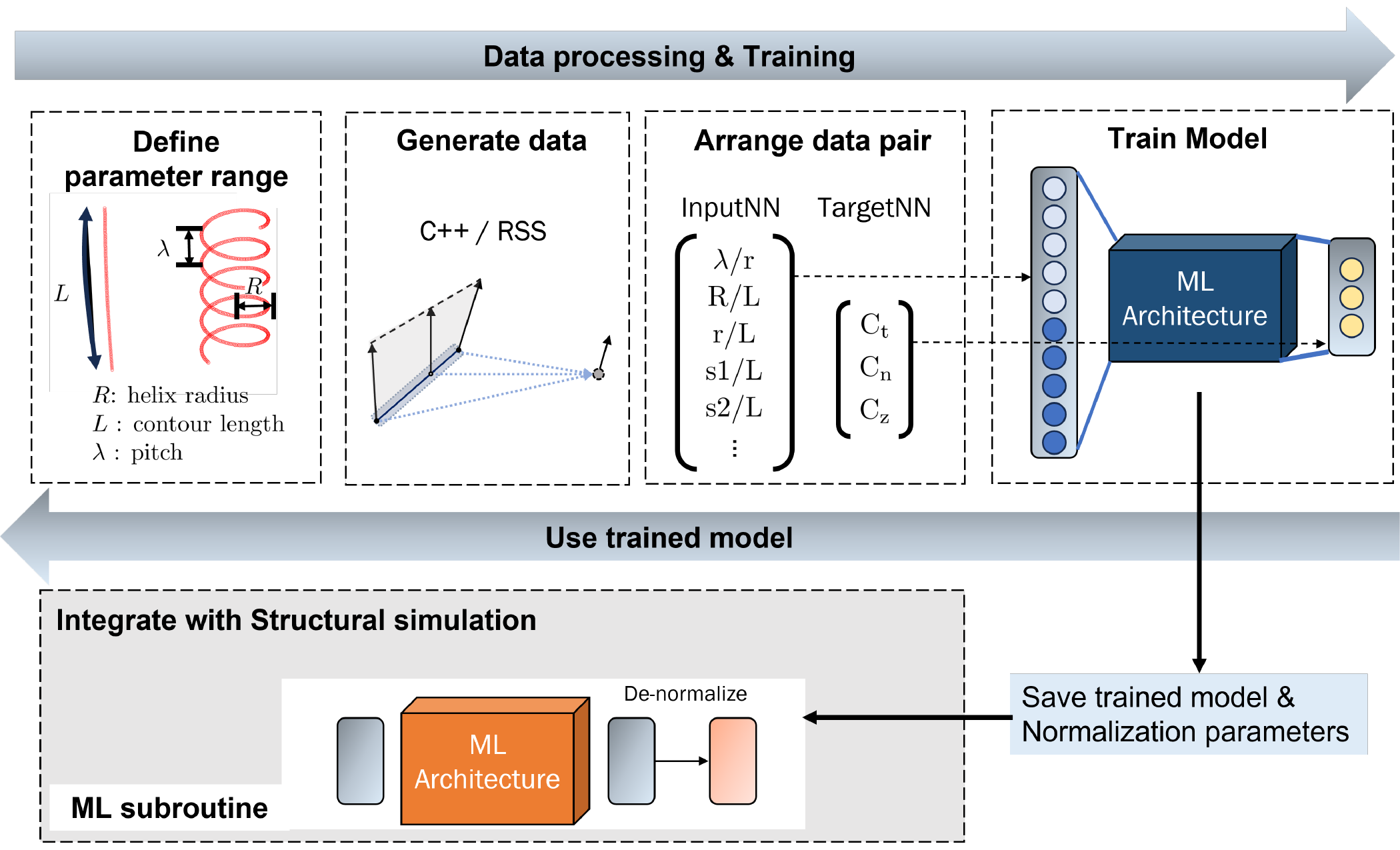}
\end{graphicalabstract}

\begin{highlights}
\item Machine learning based low Reynolds number hydrodynamics formulation for rotation and translation of helical structures.

\item The trained model and the simulations are available at

\href{https://github.com/StructuresComp/MLRFT}{https://github.com/StructuresComp/MLRFT}

\item The developed framework is comparable in accuracy with the high fidelity slender body theory but faster in computational time.

\item Possible application in real-time control of helical microbots under viscous environment.
\end{highlights}


\end{frontmatter}


\section{Introduction}
\label{sec:intro}
% Needs immense change
% Microswimmer and low Reynolds number flow
% \textbf{I added some comments to the introduction section which we can take care of at later stage.
% So I do not forget, we need to add a paragraph about some recent applications in this field.
% Emphasize on the fact that a method more accurate that RFT and faster than SBT and RSS should be used to actively control robots.}
% \note{In the current state of the art modeling, RSS and SBTs are not often used for active control due to limitations in 1) complex to apply 2)  }
% \note{Sangmin made some changes}
Resistive force theory (RFT) and slender body theory (SBT) are often compared due to the obvious pros and cons of both methods ~\cite{Johnson1979-hi,Rodenborn2013-iu,Martindale2016-oz, PhysRevFluids.6.074103}. RFT pioneered the modeling capability for biological microswimmers at low Reynolds number \cite{Vogel2012-bb,TABAK2014, Jawed2015, CICCONOFRI2019}. Gray and Hancock ~\cite{Gray1955-ca}, and Lighthill~\cite{Lighthill1975-oe} provided a practical tool by finding empirical drag coefficients for the tangential and normal motions in terms of the dimensions of the slender body. This coefficient-based theory yields simple and fast hydrodynamics calculation, and therefore, it is commonly used to model motility of bacteria and to develop various in-vivo and in-vitro microbotic systems~\cite{Peyer2014-ix,Ye2014-qd,Edd2003-vq,Riley2017,medina2016cellular,oulmas2016closed}. Meanwhile, RFTs ignore long range hydrodynamics and provides limited explanation for physical behaviors of bacterial flagella such as bundling of two flagellum or buckling of the flagella~\cite{Hoa_Nguyen2014-gi,Jawed2017,HUANG2021}. 
% Moreover, the force coefficients do not account for the local distribution, but rather represent the overall force coefficients acting on the slender object.

\begin{figure*}[!ht]
\centering
\includegraphics[width=1\textwidth]{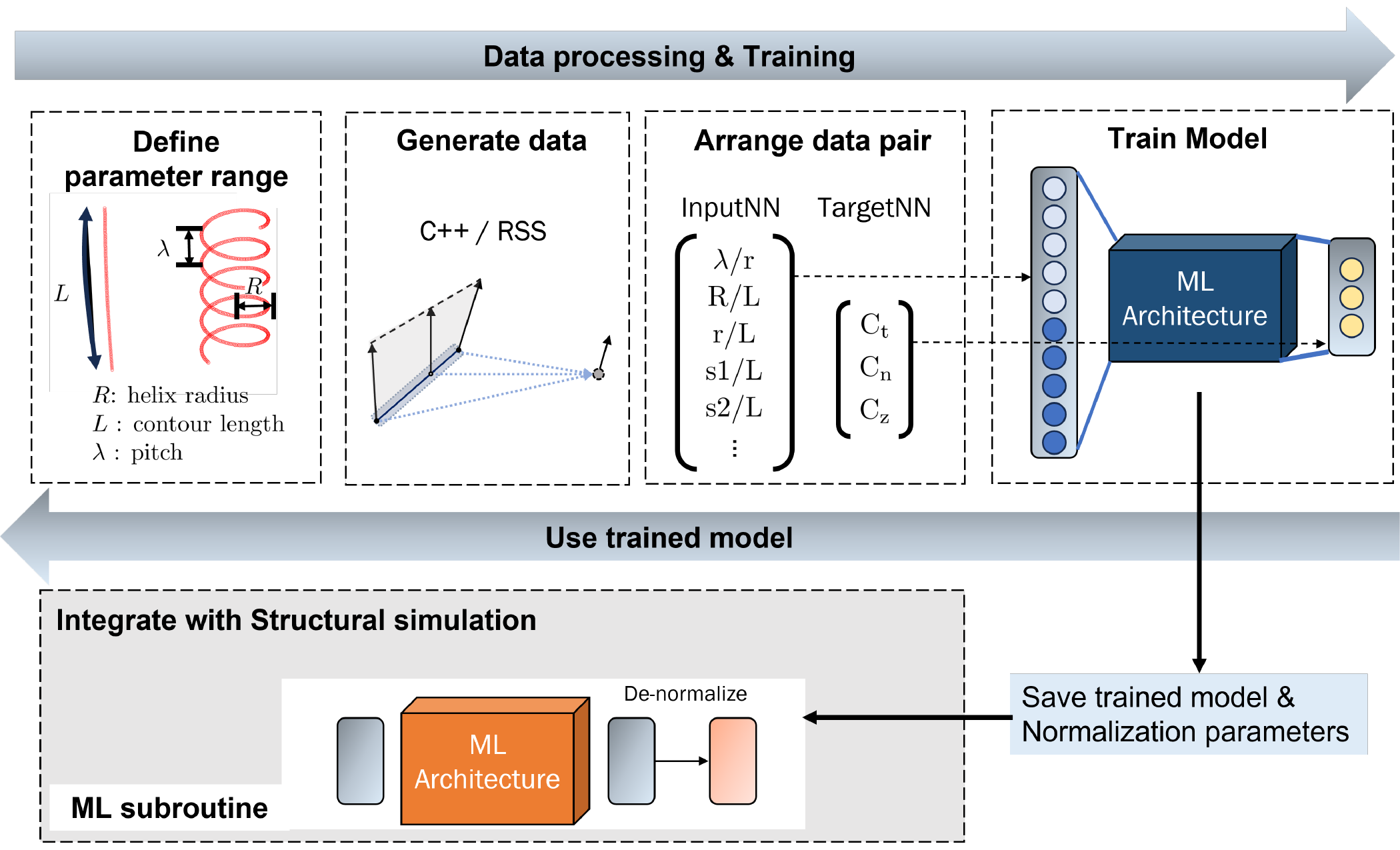}
\caption{MLRFT workflow}
\label{fig:workflow}
\end{figure*}

On the other hand, more accurate hydrodynamic method, namely the SBT, has also been used to mathematically model bacterial locomotion~\cite{SCHERR2015, Andersson2021}. However, by introducing dipoles and stokeslets, SBT associates the surface velocity of the slender body with equivalent forces exerted on the center line of the geometry. The resulting formulation demonstrates physical behavior with high accuracy due to its ability to account for the interaction of fluidic responses induced by distant parts of the flagella. Meanwhile, due to computational complexity innately present in solving a large system of linear equations, SBTs are often the limiting factor when fast computation is needed, e.g., real-time control of robotic systems.

Rodenborn \textit{et al.}~\cite{Rodenborn2013-iu} presented a robust evaluation of RFT and SBT, and quantitatively compared existing methods of RFT and SBT to experimental results on rotating and translating helical filaments. Inspired by this comparison and to exploit advantage of both methods, we delve deeper into critical evaluation of RFT with ideas to develop a new model to compensate the drawbacks of both SBT (computational complexity) and RFT (physical inaccuracy). As the first step towards the new model, we develop machine learning assisted resistive force theory (MLRFT) enabling reduced-order model that exploits advantages of each method through a simple neural network.

In this paper, our approach is to take a rotating or translating helical filament within a low Reynolds number flow. This study evaluates RFT and SBT and exploits the advantages of the two after critical evaluation of RFT against the higher-order model to formulate MLRFT. For analysis of this reduced-order model, the propulsive force, torque, and drag from MLRFT were compared against the results from an SBT to verify the accuracy. The workflow of MLRFT is described in FIG.~\ref{fig:workflow}. We begin by establishing the ranges of geometric parameters based on biological observations of bacterial flagella, including helix length, wavelength, radius, and filament radius.
% Our geometry range can also generate cilia-like slender structure to spring like helices. 
These defined parameters are then employed in a simulation based on SBT to calculate the hydrodynamic forces acting on a set of helical filaments. These filaments undergo either rotational or translational motion at low Reynolds numbers.
% Then using the geometric parameter, data are generated using the higher order hydrodynamics formulation respectively subjected to rotation and translation.
%Despite being trained on data from an SBT, a neural network is used to express drag force coefficients -- reminiscent of RFTs -- as functions of the geometry and the arc-length along the filament. For generality, the data are normalized for non-dimensional representation.
The data obtained from the SBT simulations are organized and normalized for training a neural network using the KERAS API. After the model is trained, it is saved along with the normalization values and can be integrated into structural simulations as a sub-routine to calculate external forces.
The neural network, in its current form, is restricted to helical geometries, but can be extended to include filaments of arbitrary shapes. Nonetheless, our study establishes that, instead of being restricted to a finite choice of analytical functions, neural networks can be to used to express the drag coefficients in an RFT for physical accuracy.

% \note{I modified the explanation of the workflow. I uncommented and reformulated old sentences and commented others.\\
% In the worflow replace:\\
% - add the symbols of main dimensions for the helix like lambda, L, D...\\
% - In generate data replace by a straighforward schematic like inputs\\ (lambda, L, D) [SBT]$->$ (Fx, Fy, Fz)\\
% - Arrange data pair:\\
% add some symbols in InputNN and Target NN such: InputNN(lambda, L, D) ... TargetNN(Fx, Fy, Fz)\\
% For me it seems very clear but we can try to add some more details so that the reviewer is happy.}

The rest of the paper is organized as follows. In Section~\ref{sec:evalRFT}, we discuss and introduce the assumptions behind RFTs and the formulation of MLRFT. Then in Section~\ref{sec:RSS}, a state-of-the-art SBT method, the regularized stokeslet segment (RSS), is discussed. Based on assumption and characteristics of RFT and RSS mentioned in the previous sections, Section~\ref{sec:ML} articulates the details of the data generation, neural network training, and the architecture. In Section~\ref{sec:Results}, we evaluate the performance of the MLRFT model in terms of accuracy and computational efficiency. Lastly, Section~\ref{sec:Conclusion} concludes and presents future research directions.

\begin{figure}[h!]
\centering
\includegraphics[width=0.7\textwidth]{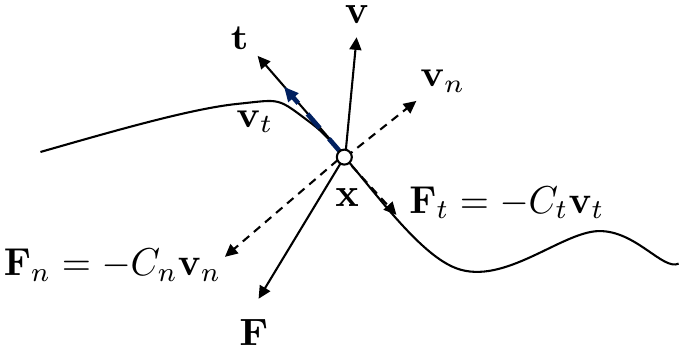}
\caption{Conceptual drawing of RFT, circled point $\mathbf x$ denotes the point of interest where we want to calculate the hydrodynamic forces.   The force at the point of interest is expressed as solid arrow stemming from the solid circle denoted as $\mathbf F$. The component of forces are divided in to velocity and tangent each represented as solid arrows denoted as $\mathbf v$ and $\mathbf t$ respectively. The components of each vector are denoted as dotted lines. From the assumption RFT only accounts for local hydrodynamic effects based on tangential and normal direction of velocity.
}
\label{fig:RFT}
\end{figure}

\section{Evaluation of RFT assumptions}
\label{sec:evalRFT}
First recall the RFT assumptions and limitations before introducing the new MLRFT method developed in this paper. RFT is the most often used hydrodynamics theory for modeling low Reynolds flow for slender structures~\cite{Darnton2007-uq,Rodenborn2013-iu,TIAN2016,Riley2017,Pozveh2021-uz,FARIS2020, Habchi2022} due to its simplicity and computational speed. RFT has proven to be practical for various applications ranging from analysis of actual bacterial flagella~\cite{Liu2011-jl,Marcos2012-dn,Bayly2011-gg,Lauga2006-kg,QIN2012} to modeling soft robots in granular medium~\cite{maladen2009undulatory, ding2012mechanics, maladen2011mechanical}. Meanwhile, RFT has several limitations due to the assumptions it is based on.

First, RFTs assume that the hydrodynamic forces can be estimated on a slender object by local coefficients. In FIG.~\ref{fig:RFT}, the hydrodynamic force per length
% \fix{Sangmin : I think this should be just force for RFT? not force per length. Khalid: Check the unit of $C_t$ and $\mathbf v_t$ and deduce the unit of $\mathbf F$.}
, $\mathbf F$, applied at the point of evaluation, $\mathbf{x}$, is only dependent upon the tangential and normal components of the velocity, namely $\mathbf{v}_t$ and $\mathbf{v}_n$, so that
\begin{equation}
    \mathbf {F} = - C_t \mathbf{v}_t - C_n \mathbf{v}_n,
    \label{eq:RFT}
\end{equation}
where $\mathbf{v}_t = (\mathbf v \cdot \mathbf t) \mathbf t$, $\mathbf v$ is the velocity of the point $\mathbf x$ with respect to the fluid, $\mathbf t$ is the tangent (unit vector) at that point, $\mathbf{v}_n = \mathbf{v} - \mathbf{v}_t$, and $C_t$ and $C_n$ are local drag coefficient to be discussed later in this section.

However, this assumption is defeated by Lighthill in 1976 \cite{Lighthill1975-oe} where he noted that constant proportionality with local velocity is inconsistent with true hydrodynamic situation and Johnson and Brokaw detailed the limitation of RFT~\cite{Johnson1979-hi} in capturing head and flagella interaction
% \fix{Lighthill does not say anything about flagella-head interaction I think : In Johnson and Brokaw (1979) We can see that lighthill has pointed out the long range interaction is important. Fixed and rephrased}
or flagella to flagella interaction. A second assumption behind RFTs is that the coefficients do not vary along the arc-length of the slender filament and ignores the long range effect of hydrodynamic interaction between distant parts of a filament. Last but not least, RFTs, e.g., the celebrated Gray and Hancock method~\cite{Gray1955-ca} and the Lighthill method~\cite{Lighthill1975-oe}, assume that the coefficients are only dependent upon the pitch to rod radius ratio, $\lambda/r$. In particular, Gray and Hancock drag coefficients are as follows \cite{Gray1955-ca}:
\begin{equation}
        \begin{aligned}
        C_t = \frac{2 \pi \mu}{\ln \frac{2\lambda}{r} - \frac{1}{2}},~
        C_n = \frac{4 \pi \mu}{\ln \frac{2\lambda}{r} + \frac{1}{2}},
    \end{aligned}
\end{equation}
whereas Lighthill RFT coefficients~\cite{Lighthill1975-oe} are
\begin{equation}
        \begin{aligned}
        C_t = \frac{2 \pi \mu}{\ln \frac{0.18\lambda}{r \cos \theta}},~
       C_n = \frac{4 \pi \mu}{\ln \frac{0.18\lambda}{r \cos \theta} +\frac{1}{2}}, \\
    \end{aligned}
\end{equation}
where $\mu$ is the viscosity, $\theta$ is pitch angle, and $C_t$ and $C_n$ represent the viscous drag coefficients along tangential and normal directions, respectively. If we want to compute the forces from velocities coupled with the structural simulation, then RFT formulation does not add extra complexity to the system, which enables high computational efficiency for this FSI solver algorithm to achieve $O(N)$ time complexity~\cite{Jawed2018,Khalid_Jawed2016-fb}. In SBTs, we have to solve an inverse problem of a dense linear system when coupled to a structural simulation. The inversion of dense matrix requires $O(N^3)$ operation and $O(N^2)$ space~\cite{Chai2011-de}, which effectively costs us computational efficiency of the FSI problem with high accuracy in return~\cite{HUANG2021,Jawed2017}. 

% Using this coefficient based methods, the major advantage that users achieve is high computational efficiency of O(N) due to only dependency on the number of the calculation points \fix{Khalid: this is not clear.  In SBT, we have to solve an inverse problem -- dense linear system and requires $O(N^3)$ operation and $O(N^2)$ space~\cite{Chai2011-de}}.

In this paper, along with the development of the MLRFT formulation, we will investigate the validity of the RFT assumption and adapt the computational advantages of RFT for the development of a fast (but physically accurate) model of hydrodynamics.

\section{Regularized Stokeslet Segments}
\label{sec:RSS}
% \continue{} 
Regularized stokeslet segment (RSS)~\cite{CORTEZ2018783} is a recently proposed SBT-like formulation of hydrodynamic forces. 
This method makes the result insensitive to spatial discretization (i.e., number of nodes on a filament) as long as the discretization level is fine enough, which is a desired trait of this type of hydrodynamic models.
% \fix{Spatial convergence} \continue{} This approach differs from the existing SBTs by involving modified force modeling which enables the exact solution to Stokes equation.
%
Cortez \textit{et al.}~\cite{CORTEZ2018783} introduced a specific regularizer,
\begin{equation}
    \phi_\epsilon({R_0}) = \frac{15\epsilon^4}{8\pi(R_0)^{7}}, ~{R_0}^2 = {|\mathbf{x}^*_i|}^2+\epsilon^2,
\end{equation}
where $\epsilon$ is a small parameter that usually is equal to the rod radius and $\mathbf{x}^*_i$ is the vector between the segment that is generating fluid flow and the point of evaluation of the hydrodynamic force. The relationship between the velocity at a point and the force per length along the slender curve of length $l$ is

\begin{figure}[h!]
\centering
\includegraphics[width=0.7\textwidth]{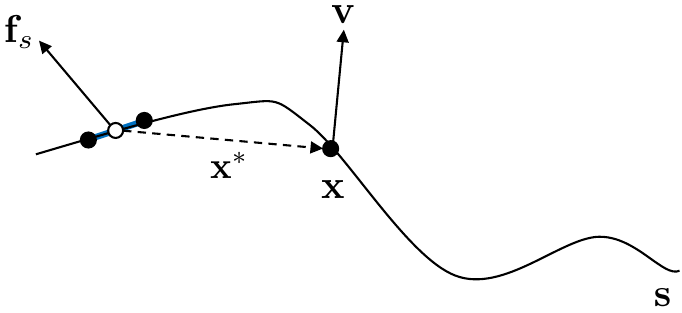}
\caption{Discretized force density description of RSS method, solid circle $\mathbf x$ denotes the point of interest where we want to calculate the hydrodynamic forces. Dotted arrow $\mathbf x^*$ represents the vector from the segment denoted in blue to the point of evaluation $\mathbf x$. Black arrow stemming from blue segment represents the discretized force density $\mathbf f_s$ of a segment along a curve $\mathbf s$. 
% \fix{We have to discuss this figure and Figure 2. Sangmin : fixed}
}
\label{fig:RSS}
\end{figure}

\begin{equation}\label{eq:RSS}
    8\pi \mu \mathbf{u(x)} =   \int_{0}^{l}
    \left[
    \left(\frac{1}{R_0}+\frac{\epsilon^2}{{R_0}^3} \right) \mathbf{f}_s + \frac{\left(\mathbf{f}_s\cdot\mathbf{x^*} \right)\mathbf{x^*}}{{R_0}^3}
    \right] \mathrm{d}s,
\end{equation}
where ${\mathbf x}$ is the point of evaluation, $\mathbf{u(x)}$ is the velocity at the point of evaluation, $l$ is a length of curve $s$ 
% \fix{I think $l$ is the length of the entire rod and NOT the line segment, Sangmin : fixed}
, $\mathbf{f}_s$ is a force per length 
% \fix{force per length? The units of Eq. 5 don't match up. N/m on LHS but N on RHS if $\mathbf f$ is ``force", Sangmin : fixed}
along the curve. Based on this fundamental, Equation \ref{eq:RSS} extends to finding out the discretized force density along a curve length based on prescribed velocity on the point of evaluation as depicted in FIG.~\ref{fig:RSS}. When the force density along a curve length is defined for each of the prescribed velocity at the point of evaluation, then the formulation makes it possible for us to find out the forces at the each point of evaluation along a line segment~\cite{CORTEZ2018783}. 
% In FIG.~\ref{fig:RSS}, $\mathbf F_i$ is \fix{Complete this sentence to describe the relationship between $\mathbf f$ and $\mathbf F_i$.}
% \fix{$\mathbf F_i$ and $\mathbf f$ -- what's the difference? :}\fix{Sangmin: $F_i$ is the sum of all the regularized forces along the curve at  where as f in the equation only refers to the force due to a line segment, should I include the discretized formulation? How to populate T and M matrix? I was more focused on delivering the essence then technicality}

Based on this formulation, RSS calculation eliminates singularity exhibited on force centered at the point of evaluation. Also, by introducing a linear continuous distribution of regularized forces along a line segment, this method decouples the values of regularization parameter from the discretization length which was a limiting factor for numerical methods. Most importantly, RSS method considers long range hydrodynamic interaction between flows induced by different discretized points on the slender structure, of which is ignored by the RFT method. FIG.~\ref{fig:RSS} describes the relationship of the non-local effect to the hydrodynamic force exerted on the point of evaluation. The dotted line between arrows represent linear continuous interpolation of the forces. Despite the advantage of being accurate and ability to account for long-range interaction within low Reynolds number flow, a major drawback of this method is the computational complexity as mentioned at the last paragraph of Section~\ref{sec:evalRFT}.
% \fix{Cite and discuss Weicheng's paper in 2 sentences. Sangmin : Weicheng's paper on bundling describes computational efficency as an advantage, added in previous section. }
The relationship established between the velocity and force on Equation~\ref{eq:RSS} shows that a dense matrix inversion is required for this long-range hydrodynamic method due to force calculation that are done based within each point of a body.

\section{Machine learning architecture and Neural network training}
\label{sec:ML}

\begin{figure*}[!ht]
\centering
\includegraphics[width=1\textwidth]{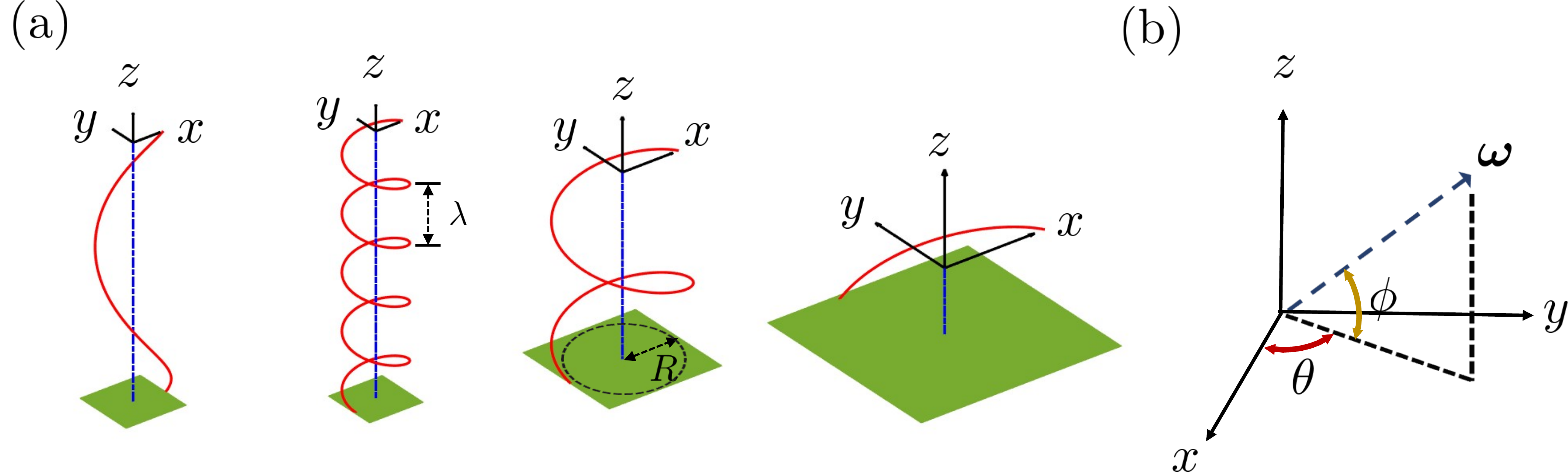}
\caption{(a) Visualized geometry data example obtained within data range of $\lambda/R$ and $L/\lambda$. Red line represents the geometry, blue dotted line represents the centerline of the helix, and the black arrows represents the axes fixed to the body frame. Pitch of helix ($\lambda$) and radius of helix ($R$) is depicted in the figure. (b) Visualized azimuth ($\theta$) and inclination angle ($\phi$) for rotational and translational velocity. The velocity vectors are defined in terms of global frame.'
% \fix{One of the helices should show $R, \lambda, \phi, \theta$., Sangmin : fixed}
}
\label{fig:geometry}
\end{figure*}

In this section, we present a detailed formulation of the MLRFT algorithm, training results, and the performance analysis of the trained model that can accurately predict the forces and torques applied on the helical structure. Based on the evaluation of the RFT done in Section~\ref{sec:evalRFT}, we hypothesized a relationship between the augmented local coefficients ($C_n, C_t, C_z$) and 10 input parameters. The augmented local coefficients function similar to the RFT coefficients,
\begin{equation}
    \mathbf F_i = \left[-C_n \left(\mathbf v_i \right)_n - C_t \left(\mathbf v_i \right)_t - C_z \left(\mathbf v_i \right)_z \right)]\mu \overline{l},
\end{equation}
 where $\mu$ is viscosity, $\overline{l}$ is voronoi length, most importantly the hydrodynamic force, $F_i$ at $i$-th node is solely dependent on the local velocity component in tangent direction,$(\mathbf v_i)_t$), normal direction,$(\mathbf v_i)_n$, and the cross of the two $(\mathbf v_i)_z$ at the point of evaluation.
The 10 input parameters were comprised of 5 geometric features and 5 velocity features that represents global/local geometry and velocity. Our goal is to train a neural network from the local force coefficients obtained through RSS and these global/local geometry/velocity input parameters.

\subsection{Data generation}
\begin{figure}[!ht]
\centering
\includegraphics[width=0.7\textwidth]{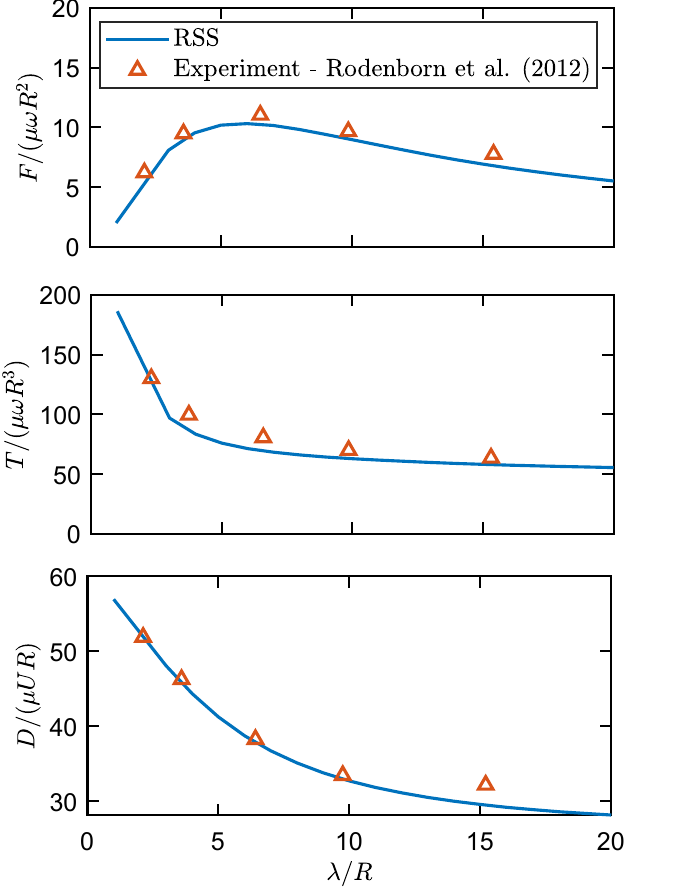}
\caption{Validation of our RSS implementation. For both experiment and simulation, a consistent ratio of geometry in length and rod radius were defined ($r=R/16$, $L=20R/\cos(\tan^{-1}(2\pi R /\lambda))$). $F,T$, and $D$ represents propulsive force, torque, and drag, respectively. The experimental values were adapted from Rodenborn et al.~\cite{Rodenborn2013-iu}}
\label{fig:rodenborn_regenerate}
\end{figure}

The range of the geometric features such as helix pitch ($\lambda$), helix radius ($R$), contour length ($L$), and rod radius ($r$) was determined based on the biological range of flagella used as a propulsive mechanism for a single cell organism. We first validate our implementation with the existing experimental data. Figure~\ref{fig:rodenborn_regenerate} validates our implementation of RSS through comparison with the experiment under same condition. The experimental values were adapted from Rodenborn et al.~\cite{Rodenborn2013-iu}.

We generated the data for the machine learning model using our RSS implementation. Table~\ref{table::Datrange} shows the geometric range in which the data were generated. The geometry space therefore amounts up to 500,000 combination for each. Some of the geometry example is depicted in FIG.~\ref{fig:geometry}(a). As shown in FIG.~\ref{fig:geometry}, the axes setup for the data generation is done in body-fixed frame. The $z$-axis is defined by helix centerline, the $x$-axis is defined by the vector with minimum distance ($R$) between the centerline and the first node of the helix, and the $y$-axis is defined by taking the cross product of $z$- and $x$-axis. The azimuth angle and inclination angle was used to determine the direction of the translation and rotational velocity in global frame, for example of rotational angle, $\omega_x =|\boldsymbol{\omega}\| \cos(\theta)\sin(\phi),~\omega_y =|\boldsymbol{\omega}\|\sin(\theta)\sin(\phi),~\omega_z =|\boldsymbol{\omega}\| \cos(\phi)$ as depicted in FIG.~\ref{fig:geometry}(b). For each geometry, the force values are calculated using RSS for each node and normalized into local coefficients that vary along the curvilinear coordinate, i.e., arclength along the filament. For each data generation case, rotational and translational flow condition is imposed.

\begin{table}[ht]
\caption{Range of helix geometry and velocity parameter for MLRFT model training. Biological bacterial flagella geometry regime is $2 < \lambda/R < 12$ , $ 2< L/\lambda < 12$ with typical filament radius of $0.01\mu m$}
\centering
\label{table::Datrange}
\begin{tabular}{p{0.4\linewidth}p{0.2\linewidth}p{0.2\linewidth}p{0.2\linewidth}}
\hline
Geometry parameter & Min. value & Max. value & Interval\\
\hline
Helix radius ($R$) & 1&1&0\\
Contour length ($L$) & 0.5$R$&30$R$&10 \\
Rod radius ($r$) & $R/100$&$R/50$&10\\
Helix pitch ($\lambda$) & 0&50$R$&50\\
%\fix{? : Explanation added on line 258 of overleaf command}
Inclination angle ($\phi$) & 0& 2$\pi$&10\\
%\fix{? : Explanation added on line 258 of overleaf command}
Azimuth angle ($\theta$) & 0& 2$\pi$&10\\
\hline
\hline
\end{tabular}
\end{table}

 After calculating over the geometric space, the geometry features were normalized to ensure scalability of the neural network model, the resultant input parameter is shown in Table~\ref{table::InputParameter}. 
 
\begin{table}[ht]
\caption{Input parameters to the DNN model}
\centering
\label{table::InputParameter}
\begin{tabular}{{ c c }}
\hline
Geometry parameter& Velocity parameter\\
\hline
$\lambda/r$ & $\|\left(\mathbf v_i \right)_t\| ~\mathtt/ ~\|\mathbf v_i\|$\\
$R/L$& $\|\left(\mathbf v_i \right)_n\| ~\mathtt/ ~\|\mathbf v_i\|$\\
$r/L$ & $\omega_x~\mathtt/ ~\|\boldsymbol{\omega}\|$ or $u_x ~\mathtt/ ~\|\mathbf u\|$\\
$s_1/L$ & $\omega_y~\mathtt/ ~\|\boldsymbol{\omega}\|$ or $u_y ~\mathtt/ ~\|\mathbf u\|$\\
$s_2/L$ & $\omega_z~\mathtt/ ~\|\boldsymbol{\omega}\|$ or $u_z ~\mathtt/ ~\|\mathbf u\|$ \\
\hline
\end{tabular}
\end{table}

% \fix{Explain normalization for $\omega$'s and $u$'s. We only need two of them (e.g., $\omega_x$ and $\omega_z$ OR $u_x$ and $u_z$). Write that normalized so that $||\omega|| = 1$ and $||u||=1$. In theory, we only need two inputs (to represent the directions) but we choose three becaues NN \ldots}\fix{Sangmin : fixed}

Here, $s_1$ represents the normalized curvilinear coordinate such that one end of the rod is $s_1=0$ and the other end is $s_1=1$, $s_2$ refers to the complementary curvilinear coordinate defined as $s_2 = 1 - s_1$,  $\|\left(\mathbf v_i \right)_t\| ~\mathtt/ ~\|\mathbf v_i\|$ and $\|\left(\mathbf v_i \right)_n\| ~\mathtt/ ~\|\mathbf v_i\|$ represent the components of the local velocity along tangential and normal directions, and $\omega_x,\omega_y,\omega_z$ represent the rotational velocity in each body-fixed direction which are imposed to the helix geometry. For the training of the translation case, $u_x,u_y,u_z$ were used instead. Both rotational and translational velocity are normalized so that $||\boldsymbol{\omega}|| = 1$ and $||\mathbf{u}||=1$. In theory, we only need two inputs for both velocity since the last directional factor can be represented by the cross product of existing ones. However, three factors were used for the NN for the robustness of our trained model. The 10 input parameters (5 geometry and 5 velocity) and 3 output coefficients $C_t$, $C_n$, and $C_z$ data pairs were each named \textbf{inputNN} and \textbf{targetNN} respectively for each node as described in FIG.~\ref{fig:workflow}. These global/local geometry and velocity features (\textbf{inputNN}) and force coefficients (\textbf{targetNN}) were used to train the neural network. 

\subsection{Machine learning architecture}

\begin{figure}[!ht]
\centering
\includegraphics[width=0.8\textwidth]{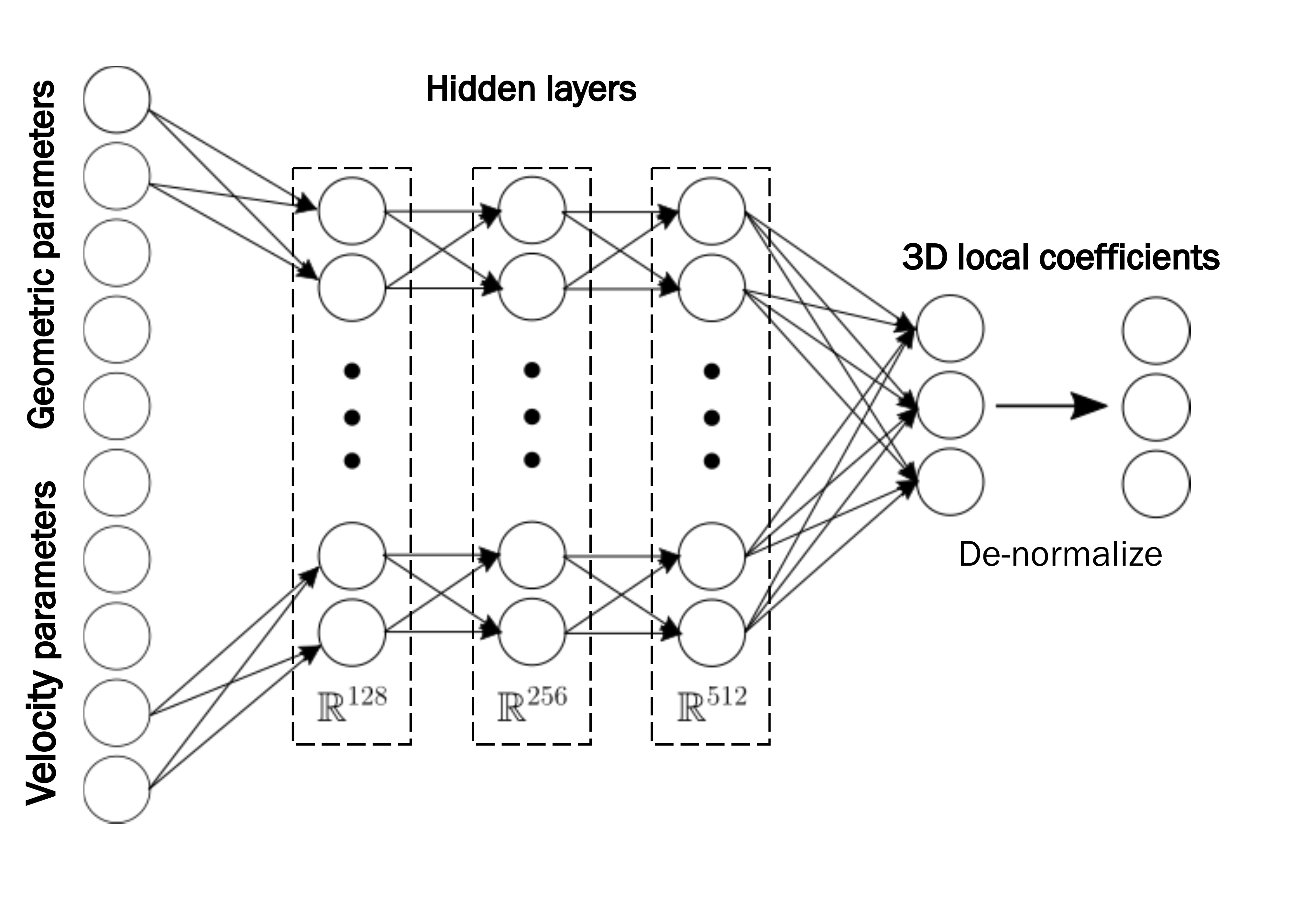}
\caption{DNN architecture of MLRFT, 10 input parameter, three hidden layers, $\mathbf{h_1}\in\mathbb{R}^{128}, \mathbf{h_2}\in\mathbb{R}^{256}, \mathbf{h_3}\in\mathbb{R}^{512}$ were used, the values of the coefficients were normalized during the training and was denormalized when used after the training.}
\label{fig:architecture}
\end{figure}

Using the data pair of input and output to the system, the relationship between the data pair were defined through artificial neural network (ANN). Despite its simplicity, the strength of ANN is that discovery of a functional relationship between the data pairs can be realized through unprecedented nonlinear pattern that were historically limited by polynomial fitting, log/exponential, and harmonic function. For our particular model, a simple multilayer perceptron (MLP) structure with three layers, each with 128, 256 and 512 neurons and Rectified Linear Unit (ReLU) activation function, were used to define the relationship between 10 inputs and 3 output coefficients. The output values were normalized across the data space in order to have matching distribution for training and test data. In a matrix form, the forward pass of first hidden layer can be represented to be

\begin{equation}
    \mathbf{h_1} = \mathbf{f}(\mathbf{W_1}\mathbf{x}+\mathbf{b_1})
\end{equation}
where $\mathbf{h_1} \in \mathbb{R}^{m\times 1}$, $\mathbf{W_1} \in \mathbb{R}^{m\times n}$, $\mathbf{x} \in \mathbb{R}^{n\times 1}$,$\mathbf{b_1} \in \mathbb{R}^{m\times 1}$ with $m = 128$ , and $n = 10$, $\mathbf{f}$ represents activation function ReLU. 

Then the relationship between each hidden layers, input and output for the neural network in FIG. \ref{fig:architecture} can be represented as:

\begin{equation}
    \mathbf{h_1} = \mathbf{f}(\mathbf{W_1}\mathbf{x}+\mathbf{b_1}),
\end{equation}
\begin{equation}
    \mathbf{h_2} = \mathbf{f}(\mathbf{W_2}\mathbf{h_1}+\mathbf{b_2}),
\end{equation}
\begin{equation}
    \mathbf{h_3} = \mathbf{f}(\mathbf{W_3}\mathbf{h_2}+\mathbf{b_3}),
\end{equation}
\begin{equation}
    \mathbf{\hat{y}} = \mathbf{f}(h_3),
\end{equation}
where $\mathbf{h_2} \in \mathbb{R}^{k\times 1}$, $\mathbf{W_2} \in \mathbb{R}^{k\times m}$, $\mathbf{b_2} \in \mathbb{R}^{k\times 1}$ with $k = 256$ and $\mathbf{h_3} \in \mathbb{R}^{l\times 1}$, $\mathbf{W_3} \in \mathbb{R}^{l\times k}$, $\mathbf{b_3} \in \mathbb{R}^{l\times 1}$ with $l = 512$. The predicted values were denoted as $\mathbf{\hat{y}}\in \mathbb{R}^{3\times 1}$ and the ground truth was represented as $\mathbf{y}\in \mathbb{R}^{3\times 1}$.

The same activation for the output layers were used in order to realize regression model. Through the training, our goal is to optimize these weights and biases for the hidden layers that are updated through gradient descent algorithm. The back propagation path works in a way by using a locally calculated gradient and then backward stepping through the optimization update. The optimization algorithm used was Adaptive moment estimation (ADAM). The training was done for 2000 epochs with learning rate of $5~\times 10^{-5}$.

The training loss function used was MAE between the predicted values and ground truth values of the normalized coefficients. The reason of choice for the loss function is due to the force coefficient peaks associated with both of ends at the geometry which is observed in RSS formulation. The normalization for the output was one of the important steps to better match the probability distribution of data across the whole data scheme. The output data set were normalized first by taking the log of the ground truth, $\mathbf{y} = \log(\mathbf{y}-\min(\mathbf{y})+2 \cdot [1,1,1]^T)$, then normalized using the mean and standard deviation for probability distribution. By taking log we could enable the values to be in the similar scale enabling faster convergence of the model. Training and test data set was divided to 70 to 30 ratio.

\begin{figure}[!ht]
\centering
\includegraphics[width=0.7\textwidth]{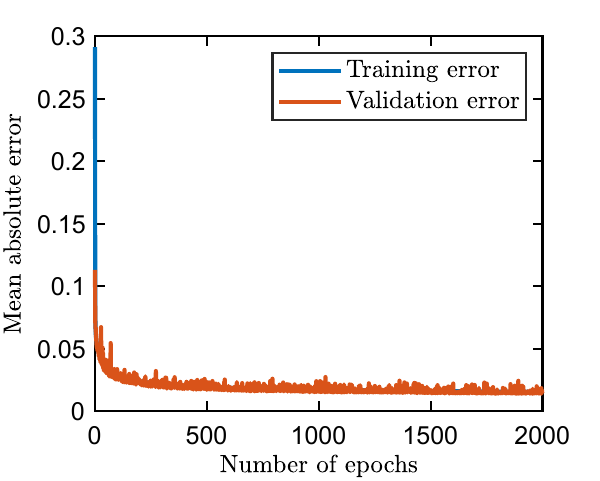}
\caption{Graph of training loss, the loss scale was chosen to be mean absolute error (MAE) in order to fit sharp jumps in forces of end nodes reported in RSS. 
%\fix{Sangmin : The loss graph is added since we are not putting graphs on appendix, as suggested by professor Habchi}
}
\label{fig:MAE loss}
\end{figure}

The training data set was divided again into training and validation data split of 80 to 20 ratio. 
%\note{CH: this sentence is not very clear} \fix{Sangmin : fixed}. 
The epoch to loss graph is depicted in FIG.~\ref{fig:MAE loss}, we can see that both training and validation loss converge throughout the epoch. When the trained model was applied for the prediction of test data, as can be shown in Table~\ref{table::Rsqrd}, we have achieved $R^2$ values for each coefficients approaching 1, which empirically shows the great match between the prediction and ground truth. 

\begin{table}[ht]
\caption{R-squared value for the coefficients}
\centering
\label{table::Rsqrd}
\begin{tabular}{{c c c c}}
\hline
 & $C_t$ & $C_n$ & $C_z$\\
\hline
$R^2$ values & 0.9958& 0.9918 & 0.9987\\
\hline
\end{tabular}
\end{table}

Throughout the training phase a computer with Intel Core i9-9920X CPU, and 4 RTX 2080 Ti GPU with RAM of 128GB was used. The training platform API was KERAS which is a subsidiary API of Tensorflow. For the execution/loading of the trained model, a computer with AMD Ryzern 7 3700x CPU, and a single NVIDIA RTX 2070 Super was used. In order to smoothly connect the structural simulation in C++ and python-trained model, we used the API called cppflow that enables model loading trained using python on C++.
% \note{Overall Algorithm of MLRFT+DER}
% \textbf{Can we move the figures showing the regression curves to this section fron the appendix or is it going to be too much information?}

\section{Results} 
\label{sec:Results}
The performance of our trained model are compared with the existing methods for the low Reynolds fluid dynamics. Owing to the objectives of this study where we would want to exploit benefit of RFT and RSS. The results are presented in two subsections. We first analyze the accuracy of newly developed MLRFT method through sweep geometry for force and torque calculation that directly relate to functionality. We then prove wrong some of the RFT assumption and compare the computational speed of the MLRFT and RSS methods. Through out the results evaluation, the Reynolds number stayed low, ($Re = 6.60\times10^{-6}<\!\!<1$ for rotation, $Re = 1.26\times10^{-5}<\!\!<1$ for translation) 
% \fix{Sangmin : Reynolds number added for the test range}

\subsection{Accuracy of MLRFT}
The accuracy of our trained model was compared and analyzed robustly. For the simulations, we coupled discrete elastic rod (DER) formulation \cite{Bergou2010,Jawed2018}
% \note{CH: to add a reference for DER were the reader can go and get more information}\fix{Sangmin : fixed} 
and the MLRFT model. The force and torque were calculated in by the external force calculation separate from the structural model. The formulation for each method for the force calculation was described in Section~\ref{sec:evalRFT} and \ref{sec:RSS}. % Add function for force and torque or at least explain it and 

\subsubsection{Force and torque comparison}
\begin{figure}[!ht]
\centering
\includegraphics[width=0.75\textwidth]{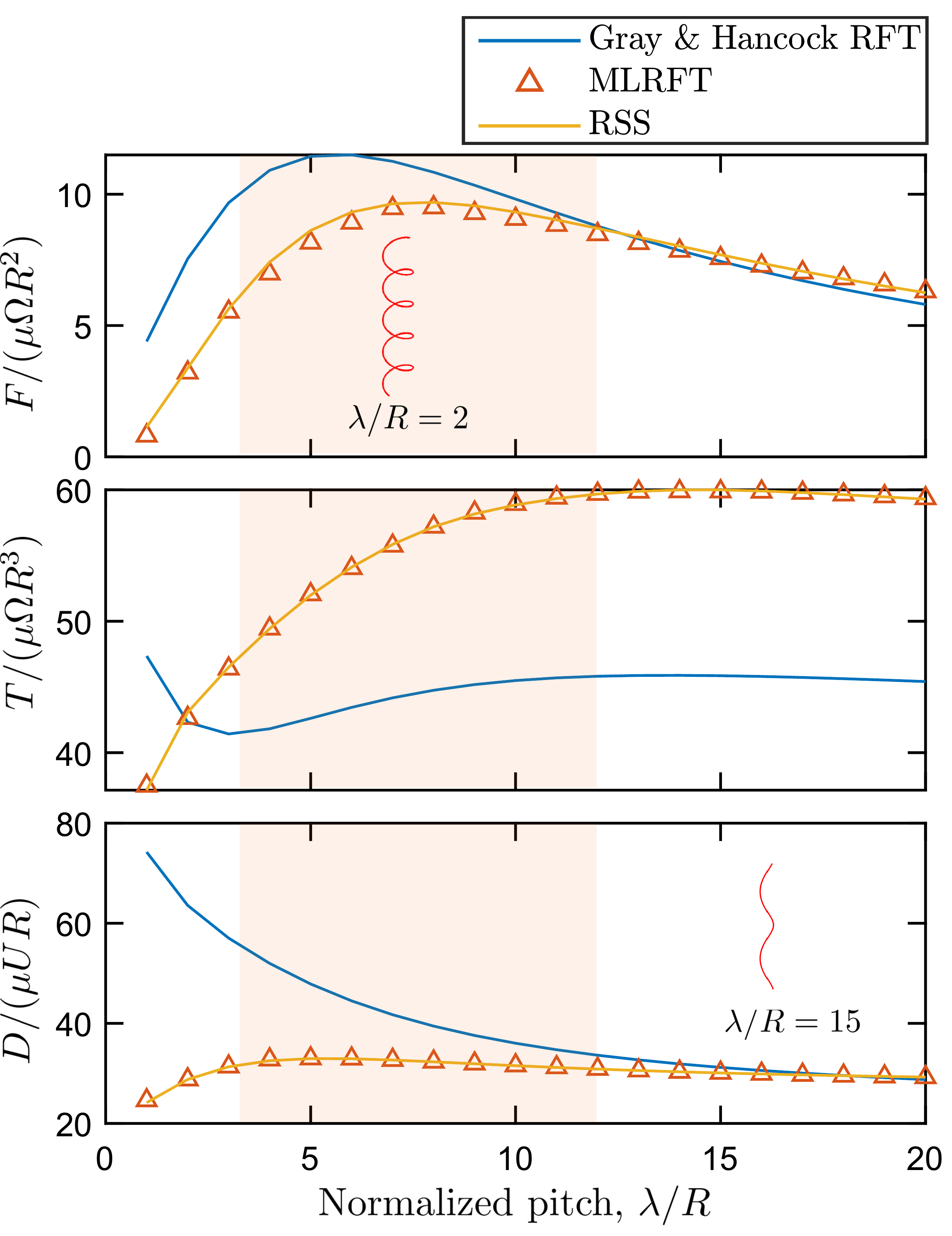}
\caption{  Normalized thrust, torque and drag of helix with independent variable $\lambda/R$. Blue solid line is the result from Gray and Hancock RFT, orange triangle is the result from MLRFT, orange solid line is the result from RSS. The rod radius, $r = R/50$ with fixed length of $L = 30R$. Shaded region represent the geometry range of bacteria found in nature.}
\label{fig:varylambda}
\end{figure}
The force and torque is one of the most crucial factor to analyze to show the hydrodynamic effect of the structure on the low Reynolds number flow because it directly relates to functionality. In this subsection, we look at the effect of geometry on the torque and force of the structure and how well can each method capture the behavior across the geometry. Here, we treat RSS method as the ground truth based on its accuracy reported on Rodenborn et. al. In FIG.~\ref{fig:varylambda}, we see the relationship between normalized pitch and the normalized force and torques. For all cases, the other geometric factors such as the rod radius and the contour length stayed the same. The trend between force and torque with the pitch shows a non linear pattern. Also, there exist optimal normalized pitch for the optimal normalized force. This shows that there exist certain design space where the effective force generation from the propulsion within low Reynolds number flow could be enabled. The machine learning based reduced order model that we trained, MLRFT has an excellent agreement with the RSS method. Also, RFT method over estimates the forces and torques in the smaller normalized pitch region and underestimates force and torque for higher pitch. The discrepancy in torque estimation was larger when compared to the RSS at a smaller normalized pitch region.

\begin{figure}[!ht]
\centering
\includegraphics[width=0.75\textwidth]{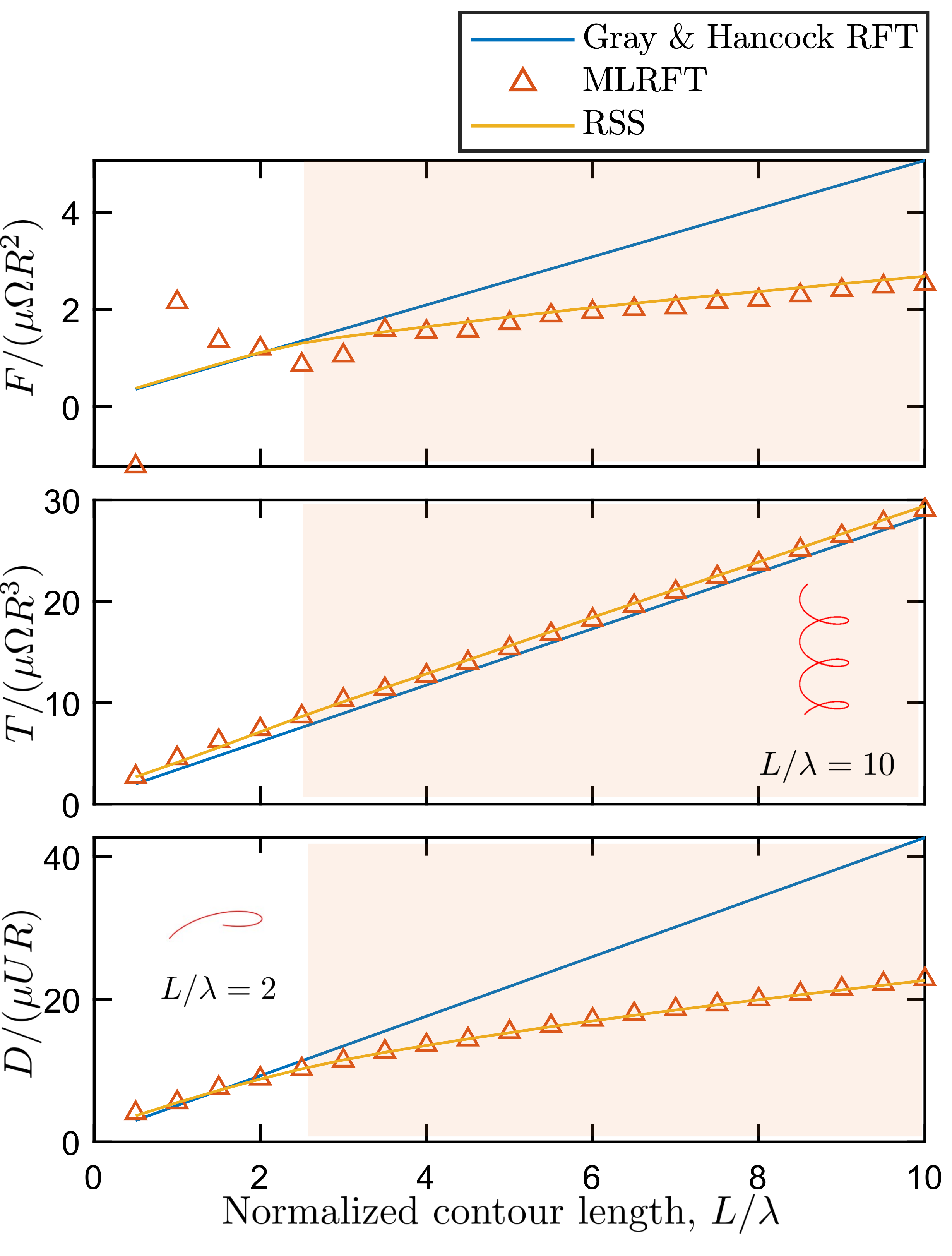}
\caption{Normalized thrust, torque and drag of helix with independent variable $L/\lambda$. Blue solid line is the result from Gray and Hancock RFT, orange triangle is the result from MLRFT, orange solid line is the result from RSS. The rod radius, $r = R/50$ with fixed lambda of $\lambda = 2R$. Shaded region represent the geometry range of bacteria found in nature.}
\label{fig:varylength}
\end{figure}

Unlike the results shown in FIG.~\ref{fig:varylambda}, the result we see in FIG.~\ref{fig:varylength} follows a linear pattern. For all the method, including RFT, RSS, and MLRFT, the resulting relationship between the length and force/torque is linear. Yet the RFT over estimates the relationship between the normalized length and the force beyond the normalized length ratio about 4. The MLRFT cannot capture the force in the low normalized contour length region due to numerical error caused by lack of discretization due to shortened length. The preset of the discrete length for the simulation when generating the force coefficients were set to be $5r$ where as the length gets smaller, the number of discretization decreases for this scheme. However, the overall performance of MLRFT follows a good trend for force and torque in the given geometric variation region.

\subsubsection{Rotational control range and acccuracy}\label{sec:rot}

\begin{figure}[!ht]
\centering
\includegraphics[width=1\textwidth]{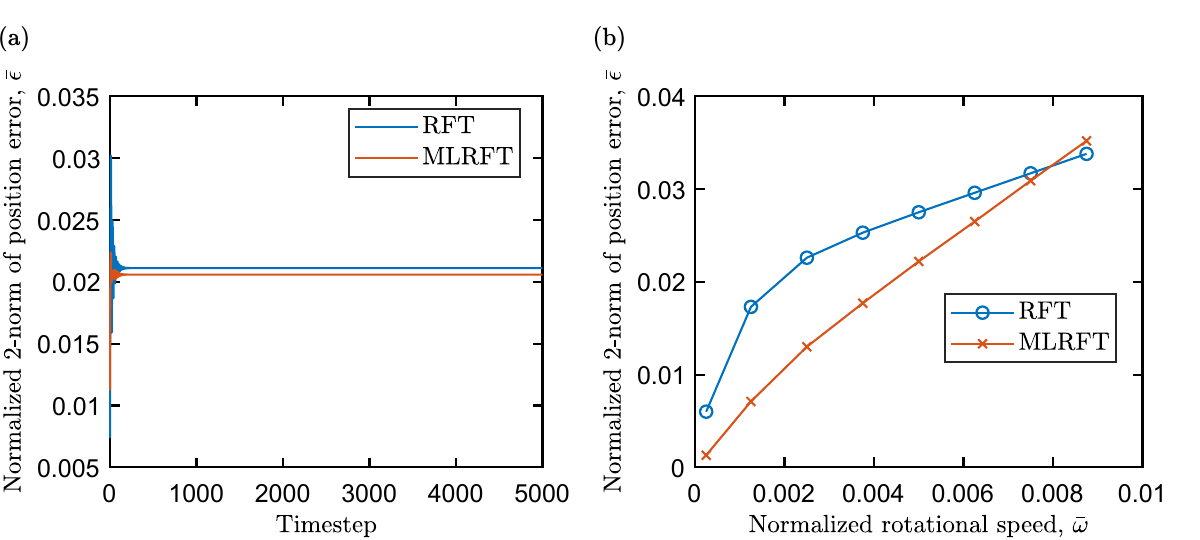}
\caption{(a) Error of RFT and MLRFT model rotating at a normalized rotational speed of $7e^{-3}$ for 5000 timesteps, 500 sec. (b) The error of RFT and MLRFT model according to the rotational speed. MLRFT shows better accuracy until the normalized rotational speed reaches $7e^{-3}$. The simulation is used for the rigid helix with geometry, $\lambda = 4R, L = 3.75\lambda, r = R/50$, and Young's modulus, $E = 100 GPa$}
\label{fig:error_rpm}
\end{figure}
To define a suitable operating range for highly accurate control of the rigid helical structure under a viscous environment, we characterized the range of rotational speed where the accuracy of our machine learning model outperforms RFT. To ensure generalizability, our results are presented in a non-dimensional format.  The normalized rotational speed is presented based on the conversion with the equation $\bar{\omega} =  \frac{\omega \mu L^4}{EI}$. The 2-norm of position error is calculated as $\bar{\epsilon}_\mathrm{RFT}=||\mathbf{x}_\mathrm{RSS} - \mathbf{x}_\mathrm{RFT}||/L$ for RFT,  and  $\bar{\epsilon}_\mathrm{MLRFT}=||\mathbf{x}_\mathrm{RSS} - \mathbf{x}_\mathrm{MLRFT}||/L$ for MLRFT. FIG.~\ref{fig:error_rpm}(b) shows that in the operating range between $0$ to $8e^{-3}$ $\bar \omega$, which corresponds to $0$ - $35$ rpm for the simulation. The MLRFT outperforms the RFT with the error magnitude twice as smaller for cases at $\bar \omega = 2e^{-3}$ and $4e^{-3}$. The error terms were obtained through the normalized Euclidean norm of position error between the RSS simulation result after 500 seconds of rotation at a single rpm. The error term converged within 100 timesteps as shown in FIG.~\ref{fig:error_rpm}(a).

\subsubsection{Validation of non-locality of MLRFT}

\begin{figure}[!ht]
\centering
\includegraphics[width=0.8\textwidth]{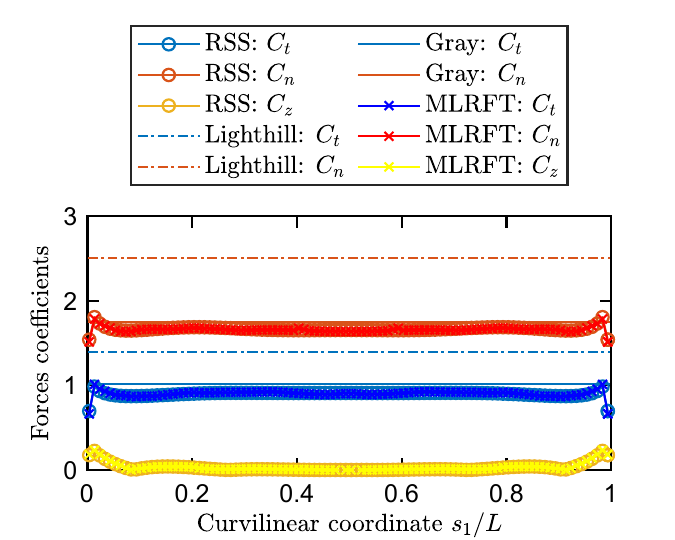}
\caption{Coefficient values for RSS method, Lighthill RFT, Gray and Hancock RFT, and MLRFT. The geometric value used was $\lambda = 8R$, $L = 3.75\lambda$, $r = R/50$. The trend shows good match with MLRFT and RSS. The values of RSS and MLRFT coefficients seem to vary with the curvilinear coordinates, while RFTs remain constant across. The Lighthill estimation is higher for both coefficients when compared to Gray and Hancock model. 
%\fix{Legend labels: ``RSS: $C_t$"}
}
\label{fig:nonlocality}
\end{figure}

As pointed out on Section~\ref{sec:evalRFT}, the RFT assumes that the coefficient variation over the curvilinear coordinates are ignored. However, we have graphed each coefficient values in FIG.~\ref{fig:nonlocality} for Gray and Hancock RFT, Lighthill RFT, RSS, and MLRFT, and found out that there exist variation in the coefficient values especially near the first, and last nodes. The result of MLRFT follows very well with the RSS. The end node (first and last nodes) are not presented due to high spikes which does not show the detail comparison of the force coefficients. However, even at the end nodes, the MLRFT provided good prediction. The results shown in the graph reevaluates the claim suggested by Johnson and Brokaw, where they claimed that the flow experienced by the flagellum is less significant without the effect of the interaction on the flow, by visualizing the similarity and the non-locality of the variation of coefficients along the curvilinear coordinates.

\subsection{Computational efficiency of MLRFT}
To calculate the computational efficiency of MLRFT, we coupled it with the discrete elastic rod (DER) method. The DER simulation is capable of enabling $O(N)$ efficiency when calculating internal elastic forces due to banded Jacobian matrix for the numerical solver process using the Newton-Raphson method. The MLRFT, RFT, and RSS methods were applied to this high-efficient simulation as an external force. Due to the fact that RSS requires a full matrix inversion process that cannot make use of banded Jacobian, RSS methods were known to have lower computational efficiency than RFT method when applied to any simulation tools. However, the RFT method can maintain the banded Jacobian structure and make sure implicit calculation possible. Using the MLRFT, we could enable highly efficient simulation with greater accuracy by eradicating the need for inverting dense matrix every step when calculating force. In order to test the computational efficiency, we varied step-size for our numerical simulation and compared it to the ratio of computational time and real time. Every geometric parameters remained the same. MLRFT follows a good efficiency trajectory as RFT. When time step is large the relative effect delay due to API relay when loading the trained model increases. The linear trend in log-log graph shows that the gap of efficiency is exponential between MLRFT and RSS.

\begin{figure}[!ht]
\centering
\includegraphics[width=0.75\textwidth]{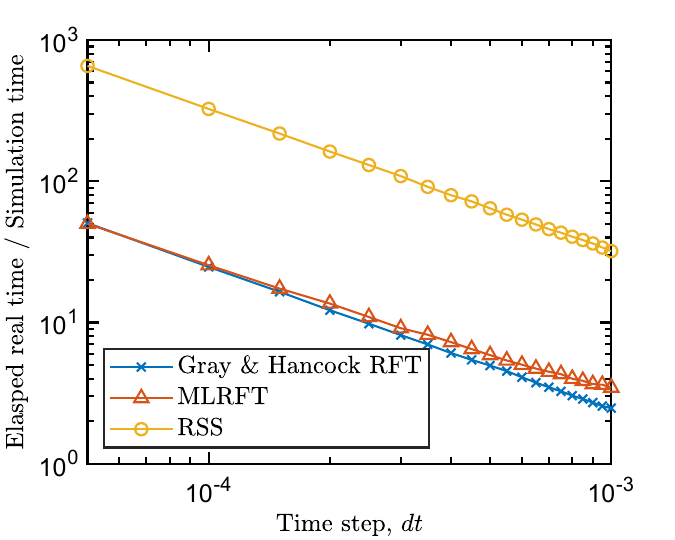}
\caption{Comparison of computational efficiency between each methods. The plot is in log scale for both axes. All simulation was run single-core, the geometric values are $R = 0.05m , \lambda = 4R, L = 7.5\lambda , \omega = 10$~rpm }
\label{fig:efficiency}
\end{figure}

% \section{Extension to non-helical objects}
% \label{sec:nonhelical}
% If we assume that a rod is piece-wise helical with pitch/radius varying as a function of the arc-length parameter, $s$, what is the improvement? One testcase is the threshold for buckling (Jawed et al. Phys. Rev. Lett.).

% \section{Deforming case analysis}
% For the future application that requires dynamic simulation of structure, we coupled the new hydrodynamic model with the structural model.
% \label{sec:deform}

\section{Concluding remarks}
\label{sec:Conclusion}
We developed a reduced-order model for low Reynolds number flow that has accuracy of an exact solution to Stokes equation for rigid slender structure using ANN. The force and torque profile for our model within geometric variation were found to The developed model also displays superiority in speed and ease of implementation. We envision this model to be applied to bacteria and cilia-inspired robots or micro-motors of which primary force/torque analysis is done through a less accurate RFT method due to complexity in implementation and low computational efficiency. Despite the high accuracy for a single/rigid geometry, our developed model is limited in accounting for the long-range interaction ability of the SBTs and assumes unbounded scenarios. We are working to develop an improved model to incorporate the elasticity, long-range hydrodynamic interaction, and boundary condition for the future. In the course of result analysis, we have discovered that the dynamic changes in long range effect characterized by current coefficients for the structure with high deformation is a topic for future investigation. The structural simulation code with the trained machine learning model is available at \href{https://github.com/StructuresComp/MLRFT}{https://github.com/StructuresComp/MLRFT}.
% \textbf{To append later.} \fix{Mention GitHub repo.} \fix{Sangmin: fixed}

\section{Acknowledgments}
We are grateful for financial support from the National Science Foundation (NSF) under award number CAREER-2047663, CMMI-2101751, and CMMI-2053971.

% \clearpage
% \appendix

% \section{Discrete elastic rod method}
% \note{CH: I believe we can remove the appendix and keep it in case reviewers ask more details.}
% This is an Appendix.
% \begin{figure}[!ht]
% \centering
% \includegraphics[width=\textwidth]{Figures/supp_fig/helix_conceptual.png}
% \caption{Discretization schematic of a helix}
% \label{fig:DER}
% \end{figure}

% \section{R-squared graph for coefficients}
% \begin{figure}[!ht]
% \centering
% \includegraphics[width=\textwidth]{Figures/supp_fig/Cn.png}
% \caption{Training result : Cn}
% \label{fig:Cn}
% \end{figure}

% \begin{figure}[!ht]
% \centering
% \includegraphics[width=\textwidth]{Figures/supp_fig/Ct.png}
% \caption{Training result : Ct}
% \label{fig:Ct}
% \end{figure}

% \begin{figure}[!ht]
% \centering
% \includegraphics[width=\textwidth]{Figures/supp_fig/Cz.png}
% \caption{Training result : Cz}
% \label{fig:Cz}
% \end{figure}

% \begin{figure}[!ht] 
% \centering
% \includegraphics[width=\textwidth]{Figures/MAELOSS.png}
% \caption{Graph of training loss, the loss scale was chosen to be mean absolute error (MAE) in order to fit sharp jumps in forces of end nodes reported in RSS. }
% \label{fig:MAE loss}
% \end{figure}

% \nocite{*}
\bibliographystyle{elsarticle-num} 
\bibliography{references}% Produces the bibliography via BibTeX.
\end{document}